# An Overview of Digital Twins Application Domains in Smart Energy Grid

Tudor Cioara, Ionut Anghel, Marcel Antal, Ioan Salomie
Claudia Antal
*Computer Science Department*
*Technical University of Cluj-Napoca*
Cluj-Napoca, Romania
{tudor.cioara, ionut.anghel, marcel.antal, ioan.salomie,
claudia.antal}@cs.utcluj.ro

Arcas Gabriel Ioan
*Engineering Powertrain Systems*
*Bosch Engineering Center*
Cluj, Romania
gabriel.arcas@ro.bosch.com

*Abstract*—**The Digital Twins (DTs) offer promising solutions for smart grid challenges related to the optimal operation, management, and control of energy assets, for safe and reliable distribution of energy. These challenges are more pressing nowadays than ever due to the large-scale adoption of distributed renewable resources at the edge of the grid. The digital twins are leveraging on technologies such as Internet of Things, big data analytics, machine learning, and cloud computing, to analyze data from different energy sensors, view and verify the status of physical energy assets and extract useful information to predict and optimize the asset's performance. In this paper, we will provide an overview of the DTs application domains in the smart grid while analyzing existing the state-of-the-art literature. We have focused on the following application domains: energy asset modeling, fault and security diagnosis, operational optimization, and business models. Most of the relevant literature approaches found are published in the last 3 years showing that the domain of DTs application in smart grid is hot and gradually developing. Anyway, there is no unified view on the DTs implementation and integration with energy management processes, thus, much work still needs to be done to understand and automatize the smart grid management.**

## I. Introduction

The increasing trend for intermittent Distributed Energy Resources (DER) adoption and deployment at the edge of the energy grid is posing serious challenges concerning the operation, management, and control for safe and reliable distribution of energy. Digital Twins (DT) offer promising solutions for these challenges, being one of the top technological trends according to Gartner [1]. The concept of DT is used to designate the virtual model of a physical asset. It was becoming increasingly popular with the digital transformation of manufacturing and distribution brought by Industry 4.0. Having a virtual model, one can apply data-driven analytics, run simulations, and what-if analyses to determine the asset behavior in real or hypothetical situations or to decide on control and optimization actions. Then the lessons learned can be applied to the physical asset manually or using actuators.

Lately, the energy ecosystem digitalization has become an important driver for achieving the targeted goals for decarbonization, cleaner energy, and efficient and secured resource management. Innovative technologies such as blockchain, or digital twins are considered the key to the transition of energy and utilities towards Energy 4.0. transition process. In particular, DTs integrate and leverage on technologies such as IoT, big data analytics, machine learning, and cloud computing, to analyze data from different sensors, view and verify the status of the physical energy asset and extract information to predict and optimize the asset's performance. They are used to virtually represent DERs and other energy assets and can be used in smart energy grid scenarios to conduct various analyses and process simulations concerning energy production, distribution, or consumption behavior [2]. Their adoption is driven by the transition process towards a sustainable and zero-carbon energy sector which aims for optimal integration and usage of renewables, lower carbon footprint, and improved energy efficiency.

In this context, the DTs allow the development of new energy services and more decentralized business models where citizens and energy resources are becoming key active players acting as prosumers and contributing to grid sustainability goals [3]. DTs can be applied for digitalizing various grid management processes such as energy production/consumption monitoring, load prediction, decision-making for energy management, balancing the supply and demand, energy security, and operation optimization [3-5]. All the above features make DTs one of the most promising emerging technology for fueling smart energy grid development and decentralization, while at the same time supporting the implementation of the next generation of energy services. Their implementation and uptake are facilitated by cloud computing which plays an important role in integrating the energy assets from electrical power systems with the data networks, offering potentially unlimited computational and storage capacity. Consequently, lately, several papers have been published on DTs applications in the smart energy grid.

In this paper, we will provide an overview of the application domains of digital twins in the smart grid while analyzing existing state-of-the-art literature. For organizing the review, we will focus on the most relevant application domains as defined by ETIP SNET [2]:

- Asset Model (DTs for energy and performance assessment and management),
- Fault Model (DTs for diagnosis of errors/faults and security issues),
- Operational Model (DTs for optimal energy distribution, energy efficiency, and cost reduction),
- Business Model (DTs for innovative energy services and new business models).

The rest of this paper is structured as follows. Section II describes a general architecture for integrating DTs in the smart grid and the technology enablers, Section III presents DTs application domains in the smart grid organized in four relevant ones, while Section IV presents the paper's conclusions.

## II. Architecture and Technology Enablers

Figure 1 shows an architecture for integrating and using DTs in smart grid applications. Several steps need to be implemented by leveraging on state-of-the-art technologies.

The first step is the creation of the DT model based on the actual measurements done on the energy assets [6]. Various characteristics can be of interest such as active of reactive power, frequency, current, energy flexibility, etc. The energy asset model is constructed by using physics or engineering models, statistics, etc. Augmented reality models can be used for improving the human understanding of the energy assets' operational states [7].

The next step addresses the coupling and integration of the DT model with the physical assets using energy sensors and actuators. The IoT sensors will be used to acquire various information regarding the energy assets operation while the actuators are used to enforce actions for control and optimization. The data monitoring will provide the software infrastructure for processing and aggregating the data received from the energy sensors.

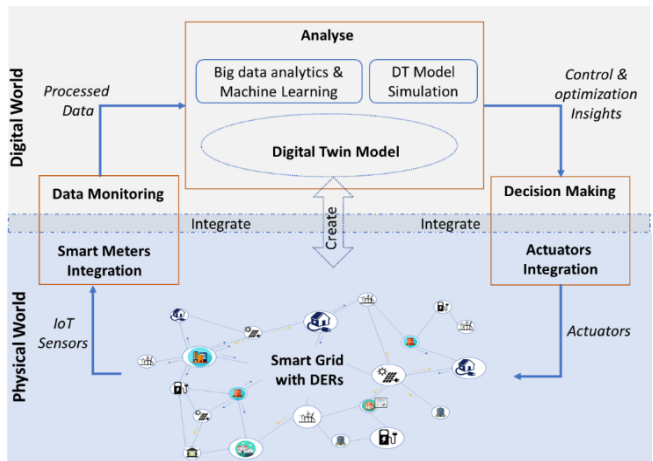

Fig. 1. DTs in smart grid applications

The third step aims to provide the technological support for running data-driven analytics on the data collected about energy assets. The smart grid can generate huge amounts of heterogeneous data thus big data analytics and machine learning are usually employed to efficiently handle them [8], [9]. Data generated by the electricity network, weather, or geographical data can be used to improve energy services delivery. The amount of data that is acquired may pose challenges regarding efficient storage and analysis using conventional approaches [10]. Additional challenges are given by the velocity at which new data is generated and moved, the variety in terms of heterogeneity of data that can be integrated, and finally the potential noise and lack of reliability of the acquisition process.

By twining data with accurate DT models of energy assets, valuable insights or information can be extracted and integrated within many smart grids' applications like operation maintenance, energy prediction, security and protection, errors detection, etc. In this context, machine learning can offer a series of benefits such as efficient load forecasting, better support for decision making or pattern detection. This kind of algorithms may be integrated with all processes and components that assure the smartness of the grid and offers better reliability in guaranteeing the quality and efficiency in energy delivery and may support the implementation of advanced services such as demand response. Machine learning allows for real-time data analysis and prediction of prosumers' energy demand or energy generation by only looking at the energy profiles. To improve accuracy and reduce model uncertainty, the energy-related features, such as demand, generation, the baseline can be fused with non-energy related vectors and contextual features determined using simulations of energy assets DTs.

Finally, the gathered insights are used for optimizing and controlling the operation of the energy assets. Model-based predictive control can be employed on both linear and non-linear systems [11], [12]. The energy assets are measured and compared to the DT model built and can be used on what if decision making to optimize the future operational states. DT models and predictive control offer promising solutions for optimizing energy efficiency, improving the performance of the energy assets, and deciding on strategies to anticipate negative events that can affect the energy grid resilience.

## III. DTs Application Domains

### A. Energy assets modeling

Energy assets modeling has an important role for smart grid operators since it allows assessing and evaluating the performance of the grid ecosystem and provides new ways for better managing the energy demand, generation, and distribution. Major industry players such as ABB [20] or Honeywell [19] have successfully approached the management of grid assets using DTs.

DTs allow the implementation of complex performance assessment models and for virtually visualize the grid resources energy behavior [3]. The development of real-time analysis of complex energy systems is proposed in [13]. The authors define a framework for DT representation by virtually modeling the grid snapshot states using DTs and applying machine learning algorithms to conduct performance assessments. In [17] the authors propose a platform for smart city energy management leveraging on DTs. Smart meters are used to gather assets data that is fed to DT virtual models of buildings with the final goal of benchmarking the building performance in terms of energy efficiency. In [25] the authors propose the development of a DT-based framework for energy demand management. The real-time data gathered from buildings is virtualized through DTs while ML models are used to offer insights to grid operators. Simulations on the DT models allow to observe demand curves and to safely manage grid response.

In other domains such as manufacturing, DTs can help to virtualize resources for achieving fine-grained energy consumption management and evaluating the energy performance of specific equipment by conducting simulations [22]. DTs are key technology enablers to build accurate models of the physical grid and carry out simulations to determine potential energy service interruptions [26]. In this way, the running of tests directly on the grid is avoided and continuous assessment platforms can be developed.

In the case of renewable energy resources, the development of DTs poses challenges such as independent monitoring of their operation to reflect their real-time functionality, the different time and scale granularity, or the intermittence of generation [16]. Mathematical models are defined to represent measurable parameters of photovoltaic systems and to further run simulations to evaluate their generation performance [14]. Similarly, wind farms and wind turbine management can be optimized by representing these assets as DTs, estimating in real-time their performance, and managing their health [15]. Characteristics and operational parameters of such assets are represented in virtual spaces and then further processed. DTs are used for representing energy storage systems virtually and assessing their performance to define operation scheduling plans before proper operation [4]. They are built upon real operation parameters of energy storage systems and simulations are used to define charging/discharging schedules based on ML techniques and decision trees.

Electric Vehicles (EVs) are growing in numbers posing serious concerns for smart grid operators. Studies highlight the added value of DTs of EV assets especially for delivering intelligent charging management. Using complex simulations, the grid can understand different types of negative scenarios such as congestion management and can avoid them by better administrating the charging stations [24]. In other approaches, DTs are used for modeling fleets of EVs and simulating their behavior to better manage their charging schedules [23]. Parameters such as energy consumption, charging capacity, and charging frequency are used to create virtual models. Time-series simulations are used for evaluating the performance of a network of EV charging stations and to make decisions for design and architecture optimization in terms of density and location.

Finally, a rather common approach for DT modeling found in the state of the art is to use semantic web and ontologies. They allow building multilayer architectures for the specific configurations of the smart grids and for various energy assets [18]. Consumers' load profiles, generation infrastructures, power supply systems can be all modeled using hierarchical structures provided by ontologies and fed as input to reasoning processes to assess their performance. Moreover, the usage of semantic web technologies for modeling energy assets allows for achieving a certain level of genericity for the modeling and simulation solutions being able to be easily adapted to different energy assets or types of smart grid architectures [21].

### B. Fault and security diagnosis

As the smart grid, operational complexity is increasing faults may occur in various components of the electrical infrastructure [38]. The main challenges are to timely diagnose and eventually avoid them. Otherwise, they can cause instabilities, energy distribution problems, losses that may culminate with power outage [37]. At the same time, cyber-attacks or natural disasters can also lead to failures, privacy breaches, and even blackouts [39]. For example, a coordinated attack on the energy grid of Ukraine has caused a power blackout that affected 225,000 households [40].

Even though DTs have promising features for addressing all these aspects for energy grids few relevant approaches can be found in the literature. In the case of dynamic systems like the smart grid, a balance should be kept between fault estimation and state estimation [6]. DTs can be used to predict potential failures, detect security issues that may affect the safe operation of energy assets, and implement mitigation actions or predictive maintenance processes [36], [37], [38]. Machine learning models and discrete-time models are proposed to detect potential warnings at the edge of the grid while transient state estimators may predict future operational faults [34]. Digital twins of grid batteries are proposed by fusing the degradation modes with machine learning techniques. This approach enables the identification and diagnosis of battery faults as well as the more advanced control of battery usage prolonging their lifetime [35]. A DT model for fault detection of drivetrains in offshore wind turbines is defined in [39]. Besides fault detection, the DTs are used to diagnose the faults and for implementing a predictive maintenance process. Mathematical models and analyses are used for fault diagnosis in the case of different energy assets. In [14] a digital twin of a Photovoltaic System energy conversion unit is described, while the fault is detected using error residual vectors, enabling real-time detection.

Even though DTs have great potential for managing more efficiently the energy assets security challenges need to be addressed [42]. Digital twins are proposed and used as a solution for analyzing potential threats to the smart grid and to assess its resilience. The DT model the grid operation to detect and avoid energy services disruptions [41]. The smart grid cybersecurity standards and potential threats are reviewed, and the authors propose the use of DTs to mitigate the lack of standardization. In [5] DT is used as a solution to mitigate coordinated attacks onto a network of interconnected power grids. Concepts like the IoT shadow and cyber-physical DT of the micro-grids network are mathematically formalized while an agent-based resilient control algorithm is implemented to detect attacks. Software-defined networks and digital twins are successfully used in [41] to improve the resilience of the smart grid in case of attacks such as distributed denial-of-service or packet delay.

### C. Grid operation and control

For optimizing energy consumption, reducing costs, and improving efficiency, DTs can act as an intermediary layer above the energy grid resources through which control decisions can be validated and enforced in real-time. Combined with trending technologies such as blockchain, they can be used to run predictions or simulations to estimate how a physical system acts in a specific situation and to improve its operational control [33]. In the case of large-scale distributed energy grids with lots of interconnected energy assets joining DTs with the decentralized nature of blockchain technology and smart contracts can further improve the smart grid operation and control starting from twinning the resources and adding them to blockchain networks [51].

DTs can pave the way for building platforms that involve complex event processing, ML, and that are mainly used to digitize the grid dispatching rules and optimize the energy distribution. DTs for Net Zero Energy Buildings are considered a solution to optimize the renewable usage and efficiency in energy consumption [27]. They deal with thermal models, energy usage, and cost analysis as DT components to optimize the energy efficiency of apartments. Simulation can be carried out also for the thermal design of a heating and cooling system of buildings using DTs [28]. For optimal energy, distribution approaches propose modeling distribution transformers as DT and computing medium voltage (MV) waveforms for identifying problems that may arise and congestion points [29]. In [30] the authors combine distributed network concepts with DT to evaluate the performance of smart grid distribution network transformers operation to defects and fix them before real-world deployment. Similarly, in [31] DTs are used to optimize energy resource usage and energy distribution paths. The model is combined with particle swarm optimization (PSO) algorithms optimization and times-series based simulations are carried out. To analyze the behavior under critical events in the power grid DT can be combined with decision-making techniques based on neural networks [32]. Such rule-events approaches can be assessed on DTs of the real-world distribution system assets to detect the critical events and to enforce intelligent control.

The complexity of the energy system is addressed using a system of system engineering as a methodology in which each sub-system integration can be optimized using DTs [52]. In [53] the authors propose an integrated energy system for testing and evaluating the impact of demand-response equipment on energy distribution. The simulation environment is modeled using a system of systems approach while hardware in the loop is used to run and control the simulation. Finally, in [54] an optimization process is implemented on top of a thermal and electrical system of system model of a DC. System-level flexibility and nonlinear optimization processes are used to grid the DC energy demand towards a target goal.

*D. Business models*

Traditionally a business model should define how an enterprise will deliver value to customers, how customers will pay for the value, and how the enterprise will convert those payments to profit [43]. In this context, DTs have been proposed for managing customer profiles, expectations and raise their awareness about the new energy services that are made available [3]. Data regarding customer preferences, satisfaction, and behavior are used to construct DTs and fed to train and run big data analytics to generate reports for customer awareness. Indeed, DTs are currently seen as a key asset for customer engagement. Using historical customer data, a DT profile can be constructed and used as a baseline for creating energy customer-targeted service offers. This kind of digital twins enabled process is applied in various domains such as transportation, online shopping, and energy grid [44], [24], [25].

In the energy domain the flexibility profiles of non-grid-owned energy assets, devices such as heat pumps, EVs, hot water boilers, are not coupled with information relating to the user's wishes in terms of comfort, convenience, and wellbeing [40]. A user's DT profile seamlessly incorporates their flexibility profiles representing the selected flexibility assets can increase the "smartness level" of buildings which will lead to increased participation in energy flexibility services [45], [17]. In this way, the small-scale energy prosumer is enabled to trade their energy flexibility and participate in flexibility-driven demand response (DR) programs. Moreover, the DTs may drive the design and implementation of cross-value chain DR business models, which enable increased consumers' DR participation and hence enable better grid management, while contributing to electricity grid reinforcement and obtaining lower energy prices for consumers [47].

In this context, the business models are based on open horizontal integration across the entire socio-economic ecosystem of the smart grid. All interested energy stakeholders will be enabled to cooperate at the digital level taking advantage of prosumers-owned energy flexibility assets. The business models will exploit digital global awareness of both energy demand and supply at the edge of the grid level facilitated by the DTs integration to define and deliver new customized energy services leveraging on increased levels of energy flexibility.

DTs facilitate the design and the development of customized energy services to appropriate clusters of prosumers organized in communities and the investigation of novel virtual aggregation business models [48], [49]. New sharing economy models can be validated in smart grid scenarios by leveraging on DTs of flexibility assets [46]. In this case, the consumers, the owners of flexible assets, and energy stakeholders will cooperate to achieve community-level welfare or energy grid sustainability objectives [47]. In that respect monetary and non-monetary incentives are split among the different local energy stakeholders, consumers and grid operators will be motivating energy consumers' participation in such energy communities. Also, the incentives can be used for additional investments in community levels energy assets for unlocking even more flexibility or for increasing the levels of renewable energy locally used [16]. In that respect, the DTs of energy assets will allow the investigation of business innovation using various configurations of shared asset ownership at the local community level.

DTs can be used for accurately estimating community-level flexibility and to find local synergies using simulations with other energy carriers such as gas, water, and heat for increasing the levels of flexibility committed [3]. Shared assets like energy storage systems lack solutions for digital integration which prevents fine-grained asset monitoring and cost-effective and efficient and reliable operation [4]. These can be facilitated by the usage of DTs and data-driven analytics to analyze the performance of the individual assets concerning performance degradation and utilization rate [22]. On top value stacking energy services, offers can be defined at the interplay among energy, mobility, health, or ambient assistive living.

Finally, the DTs can facilitate the implementation of virtual energy aggregation business models. In this case, prosumers may join in virtual coalitions such as Virtual Power Plants to deliver energy services to grid operators or to trade energy on different markets [47], [48]. The main drivers, in this case, are the increasing need to optimize the output from multiple local generation assets (i.e. wind turbines, small hydro, photovoltaic, back-up generators, etc.) and the increase of constituents' prosumers profit. The DTs can be used for analyzing and predicting the energy profiles [10] of such coalitions that mix different energy generation assets which have different energy generation models and scale with a view of reducing the associated uncertainty and optimally deliver energy services.

IV. CONCLUSIONS

In this paper, we presented an overview of the most relevant application domains of DTs in the energy field by conducting literature analysis, describing classifying the approaches found. After searching several databases, 52 papers on DT and energy have been selected and classified according to four application domains: energy asset modeling, fault and security diagnosis, grid operation optimization, and control and business models.

Even though DTs offer promising features in managing several aspects of the smart energy grid a relatively low number of papers can be found in the literature addressing the DTs applications in the energy domain. There is no consensus or unified modeling and implementation process for DTs integration with smart grid management processes. Each reviewed paper is focused on the development of specific aspects of features of DTs tailored for different energy assets, smart grid configurations, or energy services.

Anyway, since most of the literature approaches identified are from the last two or three years (over 81%) we can conclude that DTs applications in the energy grid are gradually developing the researchers starting to explore in-depth critical aspects of their implementation. But the ultimate goal of understanding and replicating completely the behavior of the physical energy asset and automating smart grid management is still far away.

ACKNOWLEDGMENT

This work has been conducted within the BRIGHT project grant number 957816 funded by the European Commission as part of the H2020 Framework Programme and it was partially supported by a grant of the Romanian Ministry of Education and Research, CNCS/CCCDI–UEFISCDI, project number PN-III-P3-3.6-H2020-2020-0031.